\documentclass[aps,prb,twocolumn,nofootinbib,notitlepage,superscriptaddress,longbibliography,preprintnumbers]{revtex4-2}
\usepackage{times}
\usepackage{graphicx}
\usepackage[justification=raggedright,singlelinecheck=false]{caption}
\usepackage{subcaption}
\usepackage[whole]{bxcjkjatype} 
\usepackage[T1]{fontenc}

\usepackage[top=30truemm,bottom=30truemm,left=25truemm,right=25truemm]{geometry}

\usepackage[colorlinks=true,linktoc=page,citecolor=red,linkcolor=blue]{hyperref}	
\graphicspath{{./figs/}}  
\usepackage[usenames]{xcolor}
\usepackage{amsmath,amssymb,amsfonts}	
\usepackage{bm,braket,array}
\usepackage{multirow}
\usepackage[all]{xy}
\usepackage{amsthm}
\usepackage{amscd}
\usepackage{slashed} 
\usepackage{calligra} 
\usepackage[normalem]{ulem} 

\theoremstyle{definition}

\theoremstyle{remark}

\newcommand{\tr}{{\rm tr\,}}

\newcommand{\pin}{{\rm Pin}}

\newcommand{\Z}{\mathbb{Z}}

\newcommand{\bk}{{\bm{k}}}

\def\widebar{\accentset{{\cc@style\underline{\mskip10mu}}}} 
\def\wideubar{\underaccent{{\cc@style\underline{\mskip10mu}}}} 

\begin{document}
\title{A Discrete Formulation of Second Stiefel-Whitney Class for Band Theory}
\author{Ken Shiozaki}
\affiliation{Center for Gravitational Physics and Quantum Information, Yukawa Institute for Theoretical Physics, Kyoto University, Kyoto 606-8502, Japan}
\author{Jing-Yuan Chen}
\affiliation{Institute for Advanced Study, Tsinghua University, Beijing, 100084, China}
\date{\today}

\begin{abstract}
Topological invariants in band theory are often formulated assuming that Bloch wave functions are smoothly defined over the Brillouin zone (BZ). However, first-principles band calculations typically provide Bloch states only at discrete points in the BZ, rendering standard continuum-based approaches inapplicable. 
In this work, we focus on the second Stiefel-Whitney class $w_2$, a key $\mathbb{Z}_2$ topological invariant under PT symmetry that characterizes various higher-order topological insulators and nodal-line semimetals. 
We develop a fully discrete, gauge-fixing-free formula for $w_2$ which depends solely on the Bloch states sampled at discrete BZ points. 
Furthermore, we clarify how our discrete construction connects to lattice field theory, providing a unifying perspective that benefits both high-energy and condensed matter approaches.
\end{abstract}
\maketitle
\parskip=\baselineskip

\section{Introduction}
Recent advances in topological band theory have greatly enhanced our understanding of electronic phases protected by various symmetries~\cite{volovik2003universe,HasanKane_RMP,QiZhang_RMP,ChiuTeoSchnyderRyu_RMP}. 
A central theme in this field is the classification of gapped (insulating) and gapless (semimetal) states by topological invariants such as Chern numbers~\cite{PhysRevLett.49.405}, the Kane--Mele $\Z_2$ invariant~\cite{PhysRevLett.95.146802,FK06}, and winding numbers~\cite{WEN1989641,PhysRevB.78.195125}. 
While these invariants are formally defined using smoothly varying Bloch wave functions (over all or part of the Brillouin zone), practical first-principles band calculations typically provide Bloch eigenstates only on a discretized BZ mesh. 
Hence, the usual continuum-based prescriptions---which rely on smooth global or piecewise gauge fixes---cannot be directly applied, posing a substantial challenge for numerical implementations.

To circumvent these difficulties, a variety of \emph{gauge-fixing-free} and \emph{discrete} formulations have been developed. 
For example, discrete algorithms exist for evaluating the first Chern number~\cite{FHS05}, and the Wilson loop approach is widely used to calculate the Kane--Mele $\mathbb{Z}_2$ invariant~\cite{PhysRevB.83.035108,PhysRevB.84.075119}. 
More recently, a discrete formula for three-dimensional winding numbers has been proposed~\cite{Shi24}.
More unified frameworks based on Wilson loop spectra~\cite{PhysRevB.93.205104}, gradient flows~\cite{tanizaki2024latticegradientflowsdestabilizing,morikawa2024windingnumber3dlattice}, and the Atiyah--Hirzebruch spectral sequence~\cite{ono2024completecharacterizationtopologicalinsulators} have also shown promise.

The second Stiefel--Whitney (SW) class $w_2$ emerges as a $\mathbb{Z}_2$ topological invariant in band structures with parity-time (PT) symmetry or the combined operation of a twofold rotation and time-reversal, characterizing both bulk properties and stable nodal structures. 
Notably, $w_2$ is pivotal in describing higher-order topological insulators and nodal-line semimetals, which are not captured by simpler indices like the first Chern number~\cite{Fang_2016,Ahn_2019}. 
A homotopy classification for a fixed number of bands can also be developed~\cite{PhysRevB.109.165125}. 
To compute $w_2$ in practice, a useful strategy is to count the crossings in Wilson loop spectra (see the Supplementary Material of Ref.~\cite{PhysRevLett.121.106403} for a review). 
Although this approach is based on a discretized BZ and does not require global gauge fixing, it relies on counting spectral crossings from a pointwise spectrum, potentially causing numerical instability near phase transitions. 
Therefore, as a complement, an alternative fully discrete, gauge-fixing-free prescription would be desirable.

In this paper, we present a fully discrete, gauge-fixing-free, and manifestly quantized algorithm for computing $w_2$ directly from the Bloch states provided on a BZ mesh. 
Our construction proceeds by diagonalizing the Bloch Hamiltonian at each discretized $\bk$ point to obtain the occupied states, forming discrete Wilson loops over each plaquette, and then lifting those loops to the $\mathrm{Pin}_+$ group.
This procedure yields a manifestly quantized $\mathbb{Z}_2$ formula without requiring any ad-hoc smoothing of Bloch wave functions. 
Since the method uses only local data at each $\bk$ point, it can be implemented straightforwardly in standard band-structure calculations.

Furthermore, we elucidate connections between our discrete approach and lattice field theory. 
By drawing analogies to gauged sigma models and Villainization procedures, we provide an interpretation of our discrete SW class in a manner that merges insights from both condensed matter and high-energy communities. 

The remainder of this paper is organized as follows. 
In Sec.~\ref{sec:Lattice formula}, we review the first and second SW classes in the continuum setting and introduce our discrete construction of $w_2$. 
In Sec.~\ref{sec:Application to band theory}, we apply this method to explicit tight-binding models, illustrating how to compute $w_2$ numerically and confirming consistency with previously known results. 
Sec.~\ref{sec:Lattice field theory} then outlines the connection to lattice field theory, clarifying how our discrete formula can be interpreted in a general field-theoretic context. 
Finally, we summarize our main findings in Sec.~\ref{sec:summary}. 
\section{Lattice Formula}
\label{sec:Lattice formula}

\subsection{Second SW class from transition function in continuum}
First, we review the definition of $w_1(P), w_2(P)$, and $W_3(P)$ for an $O(r)$ bundle $P$ over a continuum manifold $X$.
We will first provide the explicit descriptions, and then introduce some more abstract perspectives.

The isomorphism classes of principal $O(r)$ bundles $P$ are classified by the homotopy equivalence classes of continuous maps from the base space $X$ to the classifying space $BO(r) = \lim_{N \to \infty} Gr(r, \mathbb{R}^N)$, where $Gr(r, \mathbb{R}^N):= O(N)/(O(r) \times O(N - r))$ is known as the real Grassmanian. \footnote{Here, the limit $\lim$ refers to the inductive limit with respect to the standard embeddings $O(N) \hookrightarrow O(N+1)$.}
Explicitly, the classifying map $X \to BO(r)$ is specified by a continuous rank-$r$ real orthogonal projection $P_x \in \mathrm{Mat}_{N\times N}(\mathbb{R}), x \in X,$ (we use the same symbol $P$ as a principal $O(r)$ bundle) such that $P_x^2 = P_x,\, P_x^\top = P_x$, for sufficiently large $N > r$.

Let $\Phi_x = (u_{1x}, \dots, u_{rx})\in \mathrm{Mat}_{N \times r}(\mathbb{R})$ be a local frame of $P_x$, i.e. an orthonormal set of eigenvectors of $P_x$ with eigenvalue 1, so that $P_x=\Phi_x \Phi_x^\top$, and $\Phi_x^\top \Phi_x=1_{r\times r}$. Note that we can change the local frame by $\Phi_x\mapsto\Phi_x V_x$ for any $V_x\in O(r)$ without changing $P_x$. From $\Phi_x$, we can define an $SO(r)$ Berry connection $A_{\mu, x}=\Phi_x^\top \partial_{x^\mu}\Phi_x$, which gauge transforms as $A_{\mu, x}\mapsto V_x^\top (A_{\mu, x}+\partial_{x^\mu})V_x$ under change of local frame. However, while $P_x$ is continuously defined over $X\ni x$, in general $\Phi_x$ and hence $A_{\mu, x}$ cannot be continuously chosen over the entire $X$, and this is when we have a non-trivial homotopy classes of the $O(r)$ bundle. So we need to treat these steps more carefully.

\begin{figure}
\begin{subfigure}{0.2\textwidth}
\includegraphics[width=\textwidth]{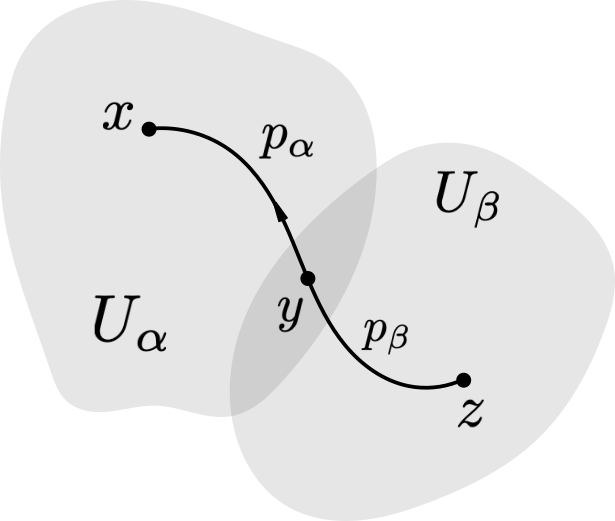}
\end{subfigure}
\hspace{1cm}
\begin{subfigure}{0.2\textwidth}
\includegraphics[width=\textwidth]{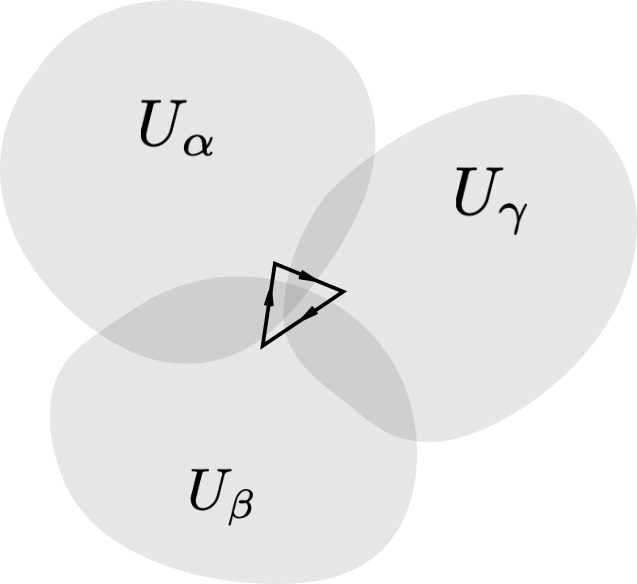}
\end{subfigure}
\caption{Left: A Wilson line crossing from $U_\beta$ to $U_\alpha$. \ Right: A tiny Wilson loop around $U_{\alpha\beta\gamma}$.}
\label{fig:Wilson_line}
\end{figure}

For a good covering $\{ U_\alpha \}_\alpha$ of $X$,\footnote{A cover $\{ U_\alpha \}_\alpha$ is called a good covering if any intersection $U_{\alpha_1 \alpha_2 \cdots}$ is contractible.} let $\Phi^\alpha_{x\in U_\alpha}$ be a local frame of $P_x$ continuously chosen over $U_\alpha$. The $SO(r)$ Berry connection $A^\alpha_{\mu, x}=(\Phi^\alpha_x)^\top \partial_{x^\mu}\Phi^\alpha_x$ is then continuously defined over $U_\alpha$.
On overlaps $U_{\alpha\beta} = U_\alpha \cap U_\beta$, transition functions $t^{\alpha\beta}_x = (\Phi^\alpha_x)^\top \Phi^\beta_x \in O(r)$ are defined, and they satisfy the cocycle condition $t^{\alpha\beta}_x t^{\beta\gamma}_x t^{\gamma\alpha}_x = 1_{r\times r}$ on triple overlaps $U_{\alpha\beta\gamma} = U_\alpha \cap U_\beta \cap U_\gamma$. 
Accordingly, $A^\beta_x=t^{\beta\alpha}_x (A^\alpha_{\mu, x}+\partial_{x^\mu}) t^{\alpha\beta}_x$.
A change in the local frame $\Phi^\alpha_x \mapsto \Phi^\alpha_x V^\alpha_x,\, V^\alpha_x \in O(r)$, induces gauge transformation in the gauge field $A^\alpha_{\mu, x}\mapsto (V^\alpha_x)^\top (A^\alpha_{\mu, x}+\partial_{x^\mu}) V^\alpha_x$ and in the transition functions $t^{\alpha\beta}_x \mapsto (V^\alpha_x)^\top t^{\alpha\beta}_x V^\beta_x$. For a path $p$ in $U_\alpha$, the Wilson line along the path is $w_p:=\mathcal{P} e^{\int_{p} A^\alpha} \in SO(r)$ \footnote{$\mathcal{P}$ means path ordering as usual.}. More generally, if $p$ crosses from $U_\beta$ to $U_\alpha$, see Fig.~\ref{fig:Wilson_line} (Left), the Wilson line is $w_p:= w_{p_\alpha} t^{\alpha\beta}_y w_{p_\beta} \in O(r)$ which satisfies $\det w_p=\det t^{\alpha\beta}_y$ and gauge transforms as $w_p\mapsto (V^\alpha_x)^\top w_p V^\beta_z$. Note $w_p$ is independent of the choice of $y\in p\cap U_{\alpha\beta}$ that divides $p$ into $p_\alpha$ and $p_\beta$.

The set of transition functions defines a \v{C}ech 1-cocycle $(-1)^{s_{\alpha\beta}} = \det t^{\alpha\beta}_x \in \{\pm 1\}$, which defines the 1st SW class $w_1(P) := [s] \in \check{H}^1(X,\Z_2)$. In terms of Wilson line, $(-1)^{s_{\alpha\beta}}=\det w_p$ for any Wilson line $w_p$ as in Fig.~\ref{fig:Wilson_line} (Left).

Given transition functions $t^{\alpha\beta}_x$, we can choose a lift $t^{\alpha\beta}_x \mapsto u(t^{\alpha\beta}_x) \in \mathrm{Pin}_+(r)$. The cocycle condition ensures that $(-1)^{z_{\alpha\beta\gamma}} := u(t^{\alpha\beta}_x) u(t^{\beta\gamma}_x) u(t^{\gamma\alpha}_x) \in \{ \pm 1 \}\subset \mathrm{Pin}_+(r)$ is a \v{C}ech 2-cocycle, and it defines the 2nd SW class  $[z] \in \check{H}^2(X, \mathbb{Z}_2)$. A change in the choice of the lift only changes $u(t^{\alpha\beta}_x)$ by $-1$, hence only changes $z$ by an exact cocycle. Therefore the cohomology class $[z]$ depends only on $t^{\alpha\beta}_x$ which defines the original bundle $P$. In terms of Wilson lines, suppose $U_{\alpha\beta\gamma}$ is a tiny region and consider three Wilson lines forming a loop around it, see Fig.~\ref{fig:Wilson_line} (Right), then $u(w_{p}) u(w_{p'}) u(w_{p''}) \rightarrow (-1)^{z'_{\alpha\beta\gamma}}$ as the loop becomes tiny compared to the spatial variation of $P_x$, with $[z']=[z] \in \check{H}^2(X, \mathbb{Z}_2)$.

These continuum Wilson line perspectives will motivate the lattice construction below.

Finally, by choosing a lift $z_{\alpha\beta\gamma} \mapsto Z_{\alpha\beta\gamma} \in \Z$, the 3rd integral SW class is given by $W_3(P) = [\delta Z/2] \in \check{H}^3(X,\Z)$, with $(\delta Z)_{\alpha\beta\gamma\delta} = Z_{\beta\gamma\delta}-Z_{\alpha\gamma\delta}+Z_{\alpha\beta\delta}-Z_{\alpha\beta\gamma}$ $\in 2\mathbb{Z}$ since $\delta z=0 \in \mathbb{Z}_2$. 
Again the cohomology class is independent of the choice of the lift.

The above are the explicit descriptions of the topological classes. More abstractly, the Stiefel-Whitney (SW) classes $w_i(P) \in H^i(X, \mathbb{Z}_2)$ are defined for $i \geq 1$.
If $X$ is finite-dimensional, $w_i(P) = 0$ for $i > \dim X$.
The SW class $w(P) = 1 + \sum_{i \geq 1} w_i(P)$ satisfies the Whitney sum formula $w(P \oplus P') = w(P)w(P')$. 
In particular, at low degrees, we have
\begin{align}
    &w_1(P \oplus P') = w_1(P) + w_1(P'), \\
    &w_2(P \oplus P') = w_2(P) + w_2(P') + w_1(P)w_1(P'). \label{eq:WSF_2}
\end{align}
The 2nd SW class $w_2(P)$ is known to be the obstruction class to having a ${\rm Pin}_+$ structure for $P$. 
Furthermore, we can write $\beta: H^2(X,\Z_2) \to H^3(X,\Z)$ for the Bockstein connecting homomorphism associated with the short exact sequence $0\to \Z \to \Z \to \Z_2 \to 0$. 
The cohomology class $W_3(P) = \beta(w_2(P)) \in H^3(X,\Z)$ is the 3rd integral SW class and is the obstruction class to having a ${\rm Pin}_c$ structure for $P$.

\subsection{Lattice second SW class}
\label{sec:Lattice second SW class}
We now construct the lattice formula for the 2nd SW class. 
We give a triangulation approximation of $X$ with a branching structure and denote the set of vertices by $\Lambda$.
At each vertex $v \in \Lambda$, we have the orthogonal projection $P_v$.
The lattice definition to be constructed will work for sufficiently fine lattice $\Lambda$, such that the variation of $P_v$ between neighboring vertices is sufficiently small. 
This lattice construction can be motivated from the continuum Wilson loop description above. 
The more systematic rationale and its relation to lattice field theory will be explained in Sec.~\ref{sec:Lattice field theory} later.

Just like in the continuum, let $\Phi_v = (u_{1v}, \dots, u_{rv})$ be a local frame of $P_v$ so that $P_v=\Phi_v \Phi_v^\top$ and $\Phi_v^\top \Phi_v=1_{r\times r}$. Since $P_v$ is a real symmetric matrix, the eigenvectors $u_{iv},\, i = 1, \dots, r$, can be chosen as real vectors.
The choice of $\Phi_v$ is not unique and has a gauge freedom $\Phi_v \mapsto \Phi_v V_v,\, V_v \in O(r)$. For an edge $(01)$, define the matrix $\tilde w_{01} := \Phi_{0}^\top \Phi_{1}$.
Under gauge transformations, it transforms accordingly
\begin{align}
    \tilde w_{01} \mapsto V_{0}^\top \tilde w_{01} V_{1}. 
    \label{eq:lattice_gauge_tr}
\end{align}
If the lattice $\Lambda$ is sufficiently fine with respect to the spatial variation of $P_x$, $\tilde w_{01}$ is close to an $O(r)$ matrix.
Performing singular value decomposition $\tilde w_{01} = L D R^\top$, we define the ``normalized" matrix
\begin{align}
    w_{01} := L R^\top \in O(r)
    \label{eq:def_w}
\end{align}
which is the lattice analog of the Wilson line. We do so for each link.
Note that $\tilde w_{10} = \tilde w_{01}^\top$ and also $w_{10} = w_{01}^\top$. 
With this, the representative simplicial 1-cocycle of the 1st SW class is given by 
\begin{align}
    (-1)^{s_{01}} = \det w_{01} \in \{\pm 1\}. 
\end{align}

For $w \in O(r)$, we define a lift $w \mapsto u(w) \in \mathrm{Pin}_+(r)$ in the following protocol. 
We can uniquely decompose $w = q_w\, R_1^{s_w}$ where $q_w\in SO(r)$, $R_1 = \mathrm{diag}(-1, 1, \dots, 1)$ is the reflection matrix of the first coordinate, and $s_w := (1-\det w)/2 \in\{0, 1\}$. Then we diagonalize $q\in SO(r)$:
\begin{align}
    q = U \begin{pmatrix}
        e^{i\theta_1} & & \\
        & \ddots & \\
        & & e^{i\theta_r}
    \end{pmatrix} U^\dag.
\end{align}
Eigenvalues $e^{i\theta_j}$ are assumed to exclude $-1$ in numerical calculations for reasons explained later.
Taking the logarithm yields
\begin{align}
    \log (q) = U \begin{pmatrix}
        i\theta_1 & & \\
        & \ddots & \\
        & & i\theta_r
    \end{pmatrix} U^\dag, \quad -\pi < \theta_j < \pi.
\end{align}
Solving the linear equation
\begin{align}
    \log (q) = \sum_{i<j} \theta_{ij} L_{ij}, 
\end{align}
with $[L_{ij}]_{kl} = -\delta_{ik} \delta_{jl} + \delta_{il} \delta_{jk}$ the generator of $SO(r)$ rotations, we get the $SO(r)$ rotation parameters $\theta_{ij}$. 
Using the gamma matrices $\gamma_1, \dots, \gamma_r$ satisfying $\{ \gamma_i, \gamma_j \} = 2\delta_{ij}$, the generators of $\mathrm{Spin}(r)$ rotations are given by $\Sigma_{ij} = \frac{[\gamma_i, \gamma_j]}{-4}$. 
We define a lift $q \mapsto u(q) \in {\rm Spin}(r)$ as 
\begin{align}
    u(q) := 
    \exp\left( \sum_{i<j} \theta_{ij} \Sigma_{ij} \right), 
\end{align}
and for $w \in O(r)$, 
\begin{align}
    u(w):= u(q_w) \times (\gamma_1)^{s_w}.
        \label{eq:def_lift}
\end{align}
In this construction 
\begin{align}
u(w^\top) = u(w)^\dag
\label{eq:transpose_dag}
\end{align}
holds.\footnote{
For $q = \exp \left[\sum_{i<j} \theta_{ij} L_{ij}\right]$, $q^\top = \exp \left[\sum_{i<j} (-\theta_{ij}) L_{ij}\right]$, meaning that $u(q)^\dag = \exp \left[\sum_{i<j} (-\theta_{ij})\Sigma_{ij}\right] = u(q^\top)$. 
For $w=q_w R_1$, $w^\top = (R_1 q_w^\top R_1) R_1$, implying $u(w^\top) = \gamma_1 u(q_w^\top) = u(w)^\dag$.}
Namely, 
\begin{align}
    u(w_{10}) = u(w_{01}^\top) = u(w_{01})^\dag
    \label{eq:lift_10_01}
\end{align}
holds.

We show why the eigenvalues of $q_w$ can be assumed not to include $-1$ for our purpose. 
The eigenvalues of an $SO(r)$ matrix $w \in SO(r)$ may have a pair of eigenvalues $\{ -1, -1 \}$. 
\footnote{One can see this by taking the logarithm of the $SO(r)$ matrix and turning the result into a tri-diagonal skew-symmetric matrix by a similarity transformation.}
The pair $\{ -1, -1 \}$ splits into pairs of complex eigenvalues $\{e^{i\theta}, e^{-i\theta}\}$ under infinitesimal perturbations.
Therefore, if $w \in SO(r)$ is random, the absence of $-1$ eigenvalues can be assumed.
The randomness of the overlap matrix $\tilde w_{01}$ can be ensured by applying random gauge transformations (\ref{eq:lattice_gauge_tr}) for all vertices or applying a small random gauge transformation when $-1$ eigenvalue appears. 
(It should be noted that to implement this procedure, the normalization $\tilde w_{01} \mapsto w_{01}$ is necessary to have the repulsion of the pair of eigenvalues $-1$.) 
A gauge transformation will lead to a change of $u(w_{10})$, including a possible jump by a $-1$ factor (especially when we use different infinitesimal gauge transformations to perturb away from the $\{ -1, -1 \}$ pair of eigenvalues in $q_w$), but as we shall see soon, the SW class will stay gauge invariant.

For a triangle $(012)$, the $O(r)$ Wilson loop 
\begin{align}
    W_{012} = w_{01}w_{12}w_{20}
\end{align}
is close to the identity matrix if the plaquette is fine enough compared to the spatial variation of $P_x$. 
Then the ${\rm Pin}_+(r)$ Wilson loop of the lifted transition functions 
\begin{align}
    {\cal W}_{012} = u(w_{01}) u(w_{12}) u(w_{20}) 
    \label{eq:pin+WL}
\end{align}
is close to either the identity $\mathbf{1}$ or $-\mathbf{1}$ as an element of the $\mathrm{Spin}(r)$ group.
We define the representative simplicial 2-cocycle of the 2nd SW class by \footnote{For example, $z_{012}=0$ when $\|{\cal W}_{012}-{\bf 1}\|<\|{\cal W}_{012}+{\bf 1}\|$, and $z_{012}=1$ otherwise. }
\begin{align}
    z_{012} :=
    \begin{cases}
        0 & \left( {\cal W}_{012} \sim {\bf 1} \right), \\
        1 & \left( {\cal W}_{012} \sim -{\bf 1} \right).
    \end{cases}
    \label{eq:z_from_W}
\end{align}
We note that the ${\rm Pin}_+$ Wilson loop (\ref{eq:pin+WL}) does not depend on the branching structure because of the relation (\ref{eq:lift_10_01}).

We show that for any closed two-dimensional surface $Y_2 \subset X$, possibly nonorientable, the signature 
\begin{align}
    (-1)^{\nu(Y_2)} = \prod_{(v_0v_1v_2) \in Y_2} (-1)^{z_{v_0v_1v_2}}
\end{align}
is indeed invariant under the gauge transformation (\ref{eq:lattice_gauge_tr}).
The Wilson line after the gauge transformation can be written as
\begin{align}
    w'_{01} = V_{0}^\top w_{01} V_{1}.
\end{align}
The lift $u(w'_{01})$ of $w'_{01}$ is independently defined from $u(w_{01})$ via (\ref{eq:def_lift}). 
However, it satisfies the following relation using the lift $u(V_v)$ of the gauge transformation matrices:
\begin{align}
    u(w'_{01}) 
    = (-1)^{\eta_{01}} u(V_{0})^\dag u(w_{01}) u(V_{1}),
\end{align}
where $\eta_{01} \in \{0,1\}$ is some 1-cochain.
Thus, the transformation of the $\mathbb{Z}_2$-valued $z_{012}$ under the gauge transformation can be written as
\begin{align}
    &z'_{012} := z_{012} + (\delta \eta)_{012}\  \bmod{2}, \\
    &(\delta \eta)_{012} := \eta_{12}+\eta_{02}+\eta_{01}\ \bmod{2}.
\end{align}
The change $(\delta \eta)_{012}$ is canceled between adjacent triangles, proving the gauge invariance of $(-1)^{\nu(Y_2)}$.
Similarly, it can be shown that $(-1)^{\nu(Y_2)}$ is invariant under the change of lift $u(w_{01}) \mapsto u(w_{01}) (-1)^{\eta_{01}}$ with arbitrary 1-cochain $\eta_{01} \in \{0,1\}$.

\subsection{Whitney Sum Formula}

Next, we demonstrate the Whitney sum formula (\ref{eq:WSF_2}).
Given two orthogonal matrices $w \in O(r)$ and $w' \in O(r')$, with their respective $SO$ rotation parameters $\theta_{ij} (1\leq i<j\leq r)$ and $\theta'_{ij} (1\leq i<j\leq r')$, we can choose the lift of their direct sum $w \oplus w'$ as $\tilde u(w\oplus w') := u(w)u'(w') \in {\rm Pin}_+(r+r')$.
Here, $u(w) = \exp\left[\sum_{i<j}\theta_{ij}\Sigma_{ij}\right] (\gamma_1)^{s_w}$, and $u'(w') = \exp\left[\sum_{i<j}\theta'_{ij}\Sigma_{i+r,j+r}\right] (\gamma_{r+1})^{s_{w'}}$.
In this choice, when both $w$ and $w'$ are odd orthogonal matrices (i.e., $\det w = \det w' = -1$), their lifts anticommute, that is,
\begin{align}
u(w)u'(w') = (-1)^{s_ws_{w'}}u'(w')u(w). 
\label{eq:uu_anticommute}
\end{align}
\footnote{From (\ref{eq:uu_anticommute}), we have $\tilde u\big((w\oplus w')^\top\big) = (-1)^{s_ws_{w'}}\tilde u(w \oplus w')^\dag$, meaning that the lift of the direct sum acquires a sign $(-1)^{s_ws_{w'}}$ under transpose, and the lift $\tilde u$ does not satisfy the relation (\ref{eq:transpose_dag}). Therefore, the lift $\tilde u(w_{ij}\oplus w'_{ij})$ depends on the branching structure.}
When the lift is defined via $\tilde u$, the ${\rm Pin}(r+r')$ Wilson loop on the triangle $(012)$, $\tilde {\cal W}_{012}:=\tilde u(w_{01}\oplus w'_{01})\tilde u(w_{12}\oplus w'_{12})\tilde u(w_{02}\oplus w'_{02})^\dag$, can be written, via straightforward calculation, as
\begin{align}
&\tilde {\cal W}_{012} = \\ &(-1)^{s'_{01}s_{12}} u(w_{01})u(w_{12}) u'(w'_{01})u'(w'_{12}) u'(w'_{20})^\dag u(w_{02})^\dag.  \nonumber
\end{align}
Here, $s_{ij}=s_{w_{ij}}, s'_{ij}=s_{w'_{ij}}$.
Noting that $u'(w'_{01})u'(w'_{12})u'(w'_{20})\sim (-1)^{z'_{012}} {\bf 1}$ and $u(w_{01})u(w_{12})u(w_{20})\sim (-1)^{z_{012}} {\bf 1}$, we finally obtain
\begin{align}
    \tilde {\cal W}_{012} \sim (-1)^{z_{012}+z'_{012}+(s'\cup s)_{012}}.
\end{align}
Taking the cohomology classes, the right-hand side is nothing but the relation (\ref{eq:WSF_2}).

\subsection{Third integral SW class}

Finally, the 3rd integral SW class $[\delta Z/2]$ is obtained by fixing one lift $z_{012} \mapsto Z_{012} \in \Z$, and for the 3-simplex $(0123)$, $\delta Z$ is given by $(\delta Z)_{0123} = Z_{123}-Z_{023}-Z_{013}+Z_{012}$. When the lattice is sufficiently fine, the $z$ constructed above will satisfy $(\delta z)_{0123}=0\in \mathbb{Z}_2$, hence indeed $(\delta Z)_{0123}/2\in\mathbb{Z}$, but it in general is not equal to $(\delta Z')_{0123}$ for any $Z'_{012}\in \mathbb{Z}$, i.e. the class $[\delta Z/2]$ may be a non-trivial integer cohomology class.

\subsection{Cubic decomposition}

Now, concerning the integral value $(-1)^{\nu(Y_2)}$ of the 2nd SW class over a two-dimensional surface $Y_2$, due to the property (\ref{eq:lift_10_01}) satisfied by the lift (\ref{eq:def_lift}), the $\Z_2$ values $z_{012}$ are defined independently of the branching structure.
Therefore, the $\Z_2$ values can be defined similarly for decompositions other than triangles.
Specifically, for (hyper)cubic lattice decompositions of $X$, the ${\rm Pin}_+$ Wilson loop on a plaquette $\Box = (0123)$ is defined as
\begin{align}
    {\cal W}_\Box = u(w_{01})u(w_{12})u(w_{23})u(w_{30})
    \label{eq:pin+Wilson_plaquette}
\end{align}
and the $\Z_2$ value $z_\Box$ is defined similarly as
\begin{align}
    z_\Box :=
    \begin{cases}
        0 & \left( {\cal W}_\Box \sim {\bf 1} \right), \\
        1 & \left( {\cal W}_\Box \sim -{\bf 1} \right).
    \end{cases}
    \label{eq:z2_plaquette}
\end{align}
The integral value of the 2nd SW class over $Y_2$ is calculated as $(-1)^{\nu(Y_2)} = \prod_{\Box \in Y_2} (-1)^{z_\Box}$.
In the following, we use this discrete expression using the (hyper)cubic lattice approximation.

\section{Application to band theory}
\label{sec:Application to band theory}

In topological band theory, when there is a PT symmetry
\begin{align}
    H_\bk^* = H_\bk,\quad \bk \in T^2,
    \label{eq:PTsym}
\end{align}
in two-dimensional (or higher) space, a $\Z_2$ topological number given by the 2nd SW class can be defined.~\footnote{More generally, the same argument applies on any closed two-dimensional manifold with PT symmetry \eqref{eq:PTsym}, not just $T^2$.}
Here, $H_\bk$ is an $N \times N$ Hermitian matrix, continuously defined on the Brillouin zone (BZ) torus $T^2$, and satisfies the gap condition.
That is, there exists a natural number $r$ such that, when the eigenvalues of $H_\bk$ are ordered in ascending order $E_{1\bk}\leq \cdots \leq E_{N\bk}$, we have $E_{r\bk}<0$ and $E_{r+1,\bk}>0$ for any $\bk \in T^2$.
The normalized negative energy eigenstates $H_\bk u_{i\bk} = E_{i\bk} u_{i\bk}, i=1,\dots,r$, define at each point a rank $r$ frame $\Phi_\bk = (u_{1\bk},\dots,u_{r\bk})$.
In particular, due to the PT symmetry (\ref{eq:PTsym}), $\Phi_\bk$ can be taken to be real, i.e., $\Phi_\bk^*=\Phi_\bk$.

Now, we partition the BZ torus into an $L \times L$ square lattice mesh, labeling the vertices as $\bm{n} = (n_x,n_y), n_x,n_y=1,\dots,L$.
At each vertex $\bm{n}$, we compute the real frame $\Phi_{\bm{n}}$, and perform a random gauge transformation $\Phi_{\bm{n}} \mapsto \Phi_{\bm{n}} V_{\bm{n}}, V_{\bm{n}} \in O(r)$.
For each edge $(\bm{n},\bm{n}+\hat \mu), \mu \in \{x,y\}$, we define the overlap matrix $\tilde w_{\bm{n},\bm{n}+\hat \mu}=\Phi_{\bm{n}}^\top \Phi_{\bm{n}+\hat \mu}$, and normalize it via SVD to obtain the Berry connection Wilson line $w_{\bm{n},\bm{n}+\hat \mu} \in O(r)$.
We define the lift $w_{\bm{n},\bm{n}+\hat \mu} \mapsto u(w_{\bm{n},\bm{n}+\hat \mu})$ according to (\ref{eq:def_lift}).
The $\Z_2$ value associated with the plaquette $\Box = (\bm{n},\bm{n}+\hat x,\bm{n}+\hat x+\hat y,\bm{n}+\hat y)$ is defined according to (\ref{eq:pin+Wilson_plaquette}) and (\ref{eq:z2_plaquette}).
The $\Z_2$ topological number given by the 2nd SW class is given by $\nu := \sum_{\Box \in T^2} z_\Box \in \{0,1\}$.

As a model including both trivial and nontrivial phases, we consider the following $4 \times 4$ Hamiltonian:
\begin{align}
    H_\bk
    &= \sin k_x \sigma_x \otimes \sigma_0 + \sin k_y \sigma_y \otimes \sigma_y \nonumber\\
    &+ (m-\cos k_x - \cos k_y) \sigma_z \otimes \sigma_0. 
\end{align}
Here, $\sigma_\mu$ are Pauli matrices.
This model has a gap closing at $m=0,2$, and it is known that the 2nd SW number is nontrivial when $0<|m|<2$.
We have confirmed that our discrete formula correctly gives the topological invariant as follows:
\begin{align}
    \nu = \begin{cases}
        1 & (0<|m|<2), \\
        0 & ({\rm else}). 
    \end{cases}
\end{align}
Moreover, we have confirmed that our discrete formula works even when $H_\bk$ is embedded into an $8 \times 8$ matrix $\tilde H_\bk = H_\bk \oplus (\sigma_z\otimes \sigma_0)$ and a perturbation $V_\bk$, a random $8 \times 8$ real symmetric matrix, is added as $\tilde H_\bk \mapsto \tilde H_\bk + V_\bk$.

Also, to confirm the validity of the Whitney sum formula (\ref{eq:WSF_2}), we consider the following $2 \times 2$ models:
\begin{align}
    &H^{(x)}_\bk(m) = \sin k_x \sigma_x + (m-\cos k_x) \sigma_z, \\
    &H^{(y)}_\bk(m) = \sin k_y \sigma_x + (m-\cos k_y) \sigma_z. 
\end{align}
These models have a nontrivial 1st SW class in the $k_x$ and $k_y$ directions respectively when $|m|<1$.
We have confirmed that our discrete formula satisfies the Whitney sum formula for the Hamiltonian $H_\bk(m_x,m_y) = H^{(x)}_\bk(m_x) \oplus H^{(y)}_\bk(m_y)$, that is,
\begin{align}
    \nu = \begin{cases}
        1 & (|m_x|,|m_y|<1), \\
        0 & ({\rm else}). 
    \end{cases}
\end{align}

\section{Rationale of lattice construction \\ and relation to lattice field theory}
\label{sec:Lattice field theory}

Our lattice construction of SW classes is at an intuitive level parallel to the continuum definition. Of course, given $P_v$ for discrete points $v\in\Lambda\subset X$, we can never fully determine $P_x$ for all $x\in X$. Roughly speaking, we need to effectively consider the possible interpolations of lattice field into a continuum field, for which a systematic understanding is developed in \cite{Chen:2024ddr}, except in the present paper we are not thinking of the lattice as the space(time) but as a discretized BZ. Then, for a sufficiently fine lattice, we take the saddle point approximation to infer ``the most probable interpolation''.

For our present problem, we just need to employ the familiar ideas of spinon decomposition and Villainization. It is instructive to first review the refined model for lattice spin $\hat{n}_v\in S^2$ which allows the 1st Chern class to be defined \cite{sachdev1990effective}. In addition to $\hat{n}_v$, the model also contains a dynamical gauge field $e^{ia_l}\in U(1)$ on each edge $l=(vv')$ that represents the Berry connection, and its Dirac string $s_p\in\mathbb{Z}$ at each plaquette $p$. The partition function is
\begin{align}
    Z=&\left[\prod_v\int_{S^2} d^2\hat{n}_v\right] \left[\prod_l\int_{-\pi}^\pi \frac{da_l}{2\pi}\right] \left[\prod_p \sum_{s_p\in \mathbb{Z}}\right] \nonumber \\
    & e^{J\sum_l \left(\tr (u_v e^{ia_l} u_{v'}^\dagger) + c.c.\right) - \frac{1}{2g^2} \sum_p (\delta a - 2\pi s)_p^2 }
\end{align}
where the ``spinon'' $u_v$ is a lift of $\hat{n}_v$ to a normalized two-component spinor so that $P_v=u_v u_v^\dagger = (1+\hat{n}_v\cdot\vec\sigma)/2$ \cite{Rabinovici:1980dn, DiVecchia:1981eh}, $a_l\in (-\pi, \pi]$ is a lift of $e^{ia_l}\in U(1)$ to $\mathbb{R}$, and $(\delta a - 2\pi s)_p\in \mathbb{R}$ is the ``Villainized'' lattice gauge flux \cite{Einhorn:1977qv} that represents the Berry curvature. Here we are viewing the $S^2$ non-linear sigma model as a ``spinon'' $SU(2)$ non-linear sigma model with a $U(1)$ gauged, and the $U(1)$ gauge field as an $\mathbb{R}$ gauge field with a 1-form $2\pi\mathbb{Z}$ gauged \cite{Chen:2024ddr}. In three-dimensions or higher, $\delta(\delta a/2\pi+s)=\delta s$ on a cube $c$ is a monopole of Berry curvature. We can include in the action a term $-\frac{U}{2} \sum_{c} \delta s_c$ to control its fugacity. Suppose we take $U\rightarrow\infty$ so that $s$ is closed (and in two-dimensions $s$ is automatically closed). The first Chern class is then given by $[s]\in H^2(X,\mathbb{Z})$.

\begin{figure}
\includegraphics[width=0.18\textwidth]{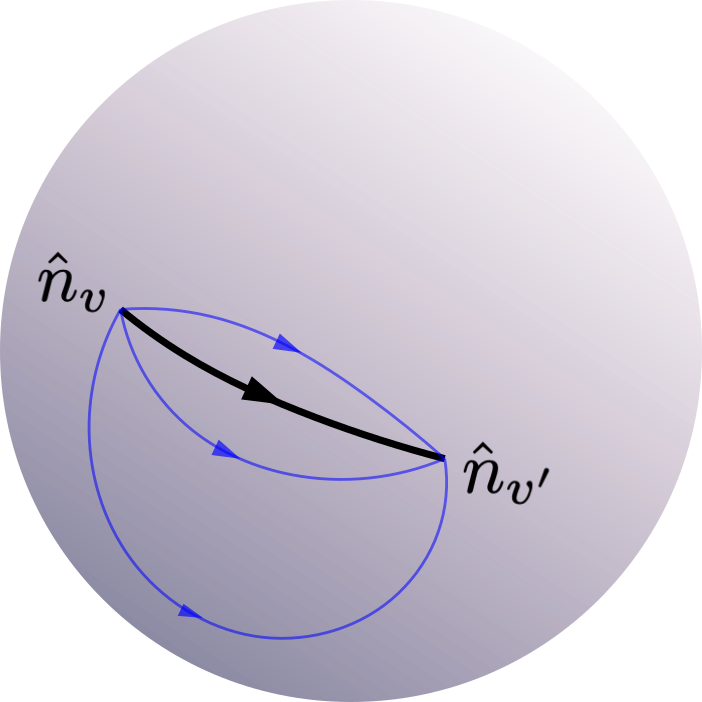}
\caption{In the refined $S^2$ non-linear sigma model, each $U(1)$ value of the lattice Berry connection Wilson line $e^{ia_l}$ corresponds to a different rotational path interpolating from $\hat{n}_v$ to $\hat{n}_{v'}$ on the target space $S^2$. Explicitly, each such rotational path is given by a $V_l\in SU(2)$ such that $V_l\hat{n}_v\cdot\vec\sigma V_l^\dagger = \hat{n}_{v'}\cdot\vec\sigma$ ($V_l$ and $-V_l$ give two opposite rotational paths on a same great circle). Such $V_l$ can be decomposed as $U_{v'}e^{-ia_l\sigma^z}U_v^\dagger$ where $U_v=(u_v, -i\sigma^y u_v^\ast) \in SU(2)$ so that $U_v\sigma^z U_v^\dagger=\hat{n}_v\cdot\vec\sigma$. The action is proportional to $-\tr V_l$ which decreases with the length of the rotational path, hence the most probable is the geodesic interpolation. See \cite{Chen:2024ddr} for more detailed discussions. Similar understanding applies to the $O(r)$ Wilson line $w_l$ that we focus on in the present paper.}
\label{fig:rot_path}
\end{figure}

In this model, for a given $\hat{n}_v\in S^2$ configuration, the Berry connection $e^{ia_l}$ and the Dirac string $s_p$ (and hence the 1st Chern class) are not determined by $\hat{n}_v$, however they are probabilistically related. How the introduction of dynamical $e^{ia_l}$ (spinon decomposition) and $s_p$ (Villainization) corresponds to considering different possible interpolations of the $S^2$ fields on $\Lambda\subset X$ into $X$ is explained in \cite{Chen:2024ddr}, and the action tells which interpolations are more probable. See Fig.~\ref{fig:rot_path} for a brief explanation. Suppose the lattice is fine enough, so that $J$ and $1/g^2$ are large, and the variation of $\hat{n}_v$ is small, then we can take the saddle point approximation, which gives $e^{ia_l}= u_v^\dagger u_{v'}/ |u_v^\dagger u_{v'}|$ (which corresponds to the familiar continuum formula $a=-iu^\dagger du$) that describes the geodesic interpolation in Fig.~\ref{fig:rot_path}, and $s_p$ equal to the integer closest to $\delta a_p/2\pi$. This indeed reduces to the method in \cite{FHS05}, see also  \cite{Berg:1981er}.

Now we can simply replace the non-linear sigma model with target space $S^2=SU(2)/U(1)$ by our Grassmanian $Gr(r, \mathbb{R}^N)=((O(N)/O(N-r))/O(r)$. The local frame $\Phi_v$ of $P_v$, analogous to the $u_v$ before, specifies an element in $O(N)/O(N-r)$, and we gauge its $O(r)$ symmetry, leading to a dynamical lattice gauge field $w_l\in O(r)$, which effectively describes how $P_v$ is interpolated along $l=(vv')$ to $P_{v'}$.
\footnote{We can also start with an $O(N)$ orthonormal frame $\Phi^{tot}=(\Phi, \Phi^\perp)$ where $\Phi^\perp$ is a local frame for the projection $P^\perp$ orthogonal to $P$. Then the $O(N-r)$ part being mod out will be acting on $\Phi^\perp$. It is familiar that working with either of $P$ or $P^\perp$ gives the same geometrical and topological information---the hidden assumption here is that the original $O(N)$ bundle over $X$ is flat and topologically trivial \cite{Moore:2017byz}, which is indeed true for the total Hilbert space over the BZ.}
The $\tr(u_v e^{ia_l} u_{v'}^\dagger)$ in the action becomes $\tr(\Phi_v w_l \Phi_{v'}^\top)$. Taking the saddle point gives our lattice formula (\ref{eq:def_w}).

The next step is to further Villainize the $O(r)$ Berry connection $w_l$ to an $\pin_+(r)$ flux $(-1)^{z_p}\mathcal{W}_p$ where $\mathcal{W}_p$ is given by (\ref{eq:pin+WL}). The $(\delta a - 2\pi s)^2$ in the action becomes $-(\tr((-1)^{z_p}\mathcal{W}_p) + c.c)$.
\footnote{Similar treatment has appeared in the Villainization of $PSU(N)$ gauge connection into $SU(N)$ flux \cite{Mack:1979gb}.}
Taking the saddle point gives our main result (\ref{eq:z_from_W}).

If we further Villainize the 2-form $\mathbb{Z}_2$ gauge field $(-1)^{z_p}$ to a 3-form $\mathbb{Z}$ flux $(\delta Z)_c - 2H_c, H_c\in\mathbb{Z}$ over each cube $c$, and prefers it to vanish at the saddle point, then we get $H=\delta Z/2$, which gives the 3rd integral SW class. In \cite{Chen:2024ddr}, in the context of $SO(3)$ connection, it is explained how this class is an obstruction to constructing a lattice ${\rm Spin}_c(3)$ connection; here this can be generalized to ${\rm Pin}_c(r)$.

\section{Summary}
\label{sec:summary}
In this work, we introduced a fully discrete, gauge-fixing-free method to compute the second SW class $w_2$ from Bloch wave functions defined on a discretized BZ. By associating the orthogonal projection of the occupied bands with discrete Wilson loops and their lifts in the ${\rm Pin}_+$ group, our approach provides a manifestly $\mathbb{Z}_2$-quantized invariant that requires neither smooth interpolation nor global gauge fixing. Hence, it is particularly well-suited for numerical band-structure calculations where Bloch states are typically obtained only on a finite momentum mesh.

We verified that our discrete formula reproduces the known $w_2$ invariants in representative models. 
In addition, we elucidated how our discrete construction connects to lattice field theory through analogies to gauged sigma models and Villainization, thereby unifying perspectives from high-energy and condensed matter physics. 

Although our focus here has been on the second SW class in PT-symmetric systems, the method can be adapted to other (magnetic) space group symmetries and superconducting systems, potentially broadening its applicability to a wider range of topological phases. 

\begin{acknowledgments}
The authors appreciate the YITP long-term workshop “Hadrons and Hadron Interactions in QCD 2024” (YITP-T-24-02) for providing
the opportunity that led to this collaboration. 
K.S. was supported by JST CREST Grant No. JPMJCR19T2, and JSPS KAKENHI Grant No. 22H05118 and 23H01097. 
J.-Y.~C. was supported by NSFC Grant No. 12342501.
\end{acknowledgments}
\vspace{4.6cm}
\pagebreak

\bibliography{ref}

\end{document}